\documentstyle[preprint,aps]{revtex}

\def\vslash{v\!\!\!\slash}

\begin{document}

{\tighten
%%\twocolumn[\hsize\textwidth\columnwidth\hsize\csname@twocolumnfalse\endcsname

\preprint{\vbox{\hbox{CALT-68-2102}
		\hbox{hep-ph/9703213}
		\hbox{\footnotesize DOE RESEARCH AND}
		\hbox{\footnotesize DEVELOPMENT REPORT} }}

\title{Model independent results for $B\to D_1(2420)\,\ell\,\bar\nu$ and \\ 
$B\to D_2^*(2460)\,\ell\,\bar\nu$ at order $\Lambda_{\rm QCD}/m_{c,b}$ }
 
\author{Adam K.\ Leibovich, Zoltan Ligeti, Iain W.\ Stewart, Mark B.\ Wise}

\address{California Institute of Technology, Pasadena, CA 91125}

\maketitle

\begin{abstract}

Exclusive semileptonic $B$ decays into $D_1$ and $D_2^*$ mesons are
investigated including order $\Lambda_{\rm QCD}/m_{c,b}$ corrections using the
heavy quark effective theory.  At zero recoil, the $\Lambda_{\rm QCD}/m_{c,b}$
corrections can be written in terms of the leading Isgur-Wise function for
these transitions, $\tau$, and known meson mass splittings.  We obtain an
almost model independent prediction for the shape of the spectrum near zero
recoil, including order $\Lambda_{\rm QCD}/m_{c,b}$ corrections.  We determine
$\tau(1)$ from the measured $B\to D_1\,\ell\,\bar\nu$ branching ratio. 
Implications for $B$ decay sum rules are discussed.

\end{abstract}

}%end tighten
\newpage 
%%\pacs{12.39.Hg, 13.20.He, 13.20.-v}
%%]\narrowtext

The use of heavy quark symmetry \cite{HQS} resulted in a dramatic improvement
in our understanding of the spectroscopy and exclusive semileptonic decays of
mesons containing a single heavy quark.  In the infinite mass limit, the spin
and parity of the heavy quark and that of the light degrees of freedom are
separately conserved.  Light degrees of freedom with quantum numbers
$s_l^{\pi_l}$ yield a doublet of meson states with total angular momentum
$J=s_l\pm\frac12$ and parity $P=\pi_l$.  All semileptonic decay form factors of
$B$ mesons into either member of such a heavy quark spin symmetry doublet are
given by just one function of $w=v\cdot v'$.  Here $v$ is the four-velocity of
the $B$ and $v'$ is that of the charmed meson.  Moreover, for the $B\to
D^{(*)}$ ground state to ground state transitions (these states have
$s_l^{\pi_l}=\frac12^-$), this universal function is normalized to unity at
zero recoil \cite{HQS,NuWe,VoSi,Luke}.  Corrections to these model independent
predictions, suppressed by powers of $\Lambda_{\rm QCD}/m_{c,b}$, can be
systematically investigated using the heavy quark effective theory (HQET)
\cite{eft}.

In this letter we discuss semileptonic $B$ meson decays into excited charmed
mesons.  Surprisingly, we find model independent predictions that hold even
including order $\Lambda_{\rm QCD}/m_{c,b}$ corrections.  We concentrate on the
doublet corresponding to $s_l^{\pi_l}=\frac32^+$, which contains the
$D_1(2420)$ and the $D_2^*(2460)$ mesons with widths around $20\,$MeV.  States
in the $s_l^{\pi_l}=\frac12^+$ doublet can decay into $D^{(*)}\pi$ in an
$s$-wave, and so they should be much broader than the $D_1$ and $D_2^*$ which
can only decay in a $d$-wave.  (An $s$-wave decay amplitude for the $D_1$ is
forbidden by heavy quark spin symmetry \cite{IWprl}.)  $B\to
D_1\,\ell\,\bar\nu$ and $B\to D_2^*\,\ell\,\bar\nu$ account for sizable
fractions of semileptonic $B$ decays \cite{ALEPH,CLEO,OPAL}, and are probably
the only three-body semileptonic $B$ decays, other than $B\to
D^{(*)}\,\ell\,\bar\nu$, whose differential decay distributions will be
precisely measured.

The measured masses of various meson states containing a bottom or charm quark
already give important information on HQET matrix elements.  The $D_2^*-D_1$
mass splitting is only $37\,$MeV, suggesting that for $s_l^{\pi_l}=\frac32^+$
states matrix elements involving the chromomagnetic operator are smaller than
for the ground state ($m_{D^*}-m_D=140\,$MeV).  The parameters $\bar\Lambda$
and $\lambda_1$ for the ground state multiplet can be related to
the analogous parameters for $D_1$ and $D_2^*$, which we denote by
$\bar\Lambda'$ and $\lambda_1'$.  We define the spin averaged masses
$\overline{m}_D=(3m_{D^*}+m_D)/4$ and
$\overline{m}_D'=(5m_{D_2^*}+3m_{D_1})/8$, and similarly for analogous mesons
containing a bottom quark.  Using the measured hadron masses \cite{PDG}, and
identifying $\overline{m}_B'=5.70\,$GeV as the mass of the $B_J^*(5732)$
state, we find
\begin{eqnarray}\label{LpL}
\lambda_1'-\lambda_1 &=& 2m_cm_b\, (\overline{m}_B' - \overline{m}_B
  - \overline{m}_D' + \overline{m}_D) / (m_b-m_c) 
  \simeq -0.34\,{\rm GeV}^2 \,, \nonumber\\*
\bar\Lambda'-\bar\Lambda &=& \overline{m}_D'-\overline{m}_D
  + (\lambda_1'-\lambda_1)/(2m_c) \simeq0.35\,{\rm GeV} \,.
\end{eqnarray}

The matrix elements of the vector and axial currents 
($V^\mu=\bar c\,\gamma^\mu\,b$ and $A^\mu=\bar c\,\gamma^\mu\gamma_5\,b$) 
between $B$ mesons and $D_1$ or $D_2^*$ mesons can be parametrized as
\begin{eqnarray}\label{formf}
\langle D_1(v',\epsilon)|\, V^\mu\, |B(v)\rangle 
  &=& \sqrt{m_{D_1}m_B}\, [ f_{V_1}\, \epsilon^{*\mu} 
  + (f_{V_2} v^\mu + f_{V_3} v'^\mu)\, (\epsilon^*\cdot v)] \,, \nonumber\\*
\langle D_1(v',\epsilon)|\, A^\mu\, |B(v)\rangle 
  &=& \sqrt{m_{D_1}m_B}\, i\, f_A\, \varepsilon^{\mu\alpha\beta\gamma} 
  \epsilon^*_\alpha v_\beta v'_\gamma \,, \nonumber\\
\langle D^*_2(v',\epsilon)|\, A^\mu\, |B(v)\rangle 
  &=& \sqrt{m_{D_2^*}m_B}\, [ k_{A_1}\, \epsilon^{*\mu\alpha} v_\alpha 
  + (k_{A_2} v^\mu + k_{A_3} v'^\mu)\,
  \epsilon^*_{\alpha\beta}\, v^\alpha v^\beta ] \,, \nonumber\\*
\langle D^*_2(v',\epsilon)|\, V^\mu\, |B(v)\rangle 
  &=& \sqrt{m_{D_2^*}m_B}\, i\,k_V\, \varepsilon^{\mu\alpha\beta\gamma} 
  \epsilon^*_{\alpha\sigma} v^\sigma v_\beta v'_\gamma \,. 
\end{eqnarray}
Here $f_i$ and $k_i$ are functions of $w$.  The differential decay rates for
$B\to D_1\,\ell\,\bar\nu$ and $B\to D_2^*\,\ell\,\bar\nu$ decays in terms of
these form factors are, respectively ($r_1=m_{D_1}/m_B$ and 
$r_2=m_{D_2^*}/m_B$)
\begin{mathletters}\label{rates}
\begin{eqnarray}
{{\rm d}\Gamma_1 \over{\rm d}w} &=& 
  {G_F^2\,|V_{cb}|^2\,m_B^5\,r_1^3 \over48\pi^3}\, \sqrt{w^2-1}\,
  \bigg\{ 2(1-2wr_1+r_1^2)\, 
  \Big[ f_{V_1}^2 + (w^2-1)\,f_A^2 \Big] \nonumber\\*
&& + \Big[ (w-r_1)\,f_{V_1} 
  + (w^2-1)\, (f_{V_3}+r_1 f_{V_2}) \Big]^2 \bigg\} \,, \\
{{\rm d}\Gamma_2 \over{\rm d}w} &=& 
  {G_F^2\,|V_{cb}|^2\,m_B^5\,r_2^3 \over144\pi^3}\, (w^2-1)^{3/2}\, 
  \bigg\{ 3(1-2wr_2+r_2^2)\, \Big[ k_{A_1}^2 + (w^2-1)\,k_V^2 \Big] \nonumber\\*
&& + 2\Big[ (w-r_2)\,k_{A_1} 
  + (w^2-1)\, (k_{A_3}+r_2\,k_{A_2}) \Big]^2 \bigg\} \,. 
\end{eqnarray}
\end{mathletters}%

The form factors $f_i$ and $k_i$ can be parametrized by a set of Isgur-Wise
functions at each order in $\Lambda_{\rm QCD}/m_{c,b}$.  It is simplest to
calculate the matrix elements in Eq.~(\ref{formf}) using the trace formalism
\cite{trace}.  The fields $P_v$ and $P_v^{*\mu}$ that destroy members of the
$s_l^{\pi_l}=\frac12^-$ doublet with four-velocity $v$ are in the $4\times4$
matrix
\begin{equation}\label{Hdef}
H_v = \frac{1+\vslash}2\, \Big[ P_v^{*\mu} \gamma_\mu 
  - P_v\, \gamma_5 \Big] \,,
\end{equation} 
while for $s_l^{\pi_l}=\frac32^+$ the fields $P_v^\nu$ and 
$P_v^{*\mu\nu}$ are in
\begin{equation}\label{Fdef}
F_v^\mu = \frac{1+\vslash}2 \bigg\{ \! P_v^{*\mu\nu} \gamma_\nu 
  - \sqrt{\frac32}\, P_v^\nu \gamma_5 \bigg[ g^\mu_\nu - 
  \frac13 \gamma_\nu (\gamma^\mu-v^\mu) \bigg] \bigg\} .
\end{equation}  
The matrices $H$ and $F$ satisfy $\vslash H_v=H_v=-H_v\vslash$,~
$\vslash F_v^\mu=F_v^\mu=-F_v^\mu\vslash$,~ $F_v^\mu\gamma_\mu=0$, and 
$v_\mu F_v^\mu=0$.  

To leading order in $\Lambda_{\rm QCD}/m_{c,b}$ and $\alpha_s$
\begin{equation}\label{lo}
\bar c\, \Gamma\, b = \bar h^{(c)}_{v'}\, \Gamma\, h^{(b)}_v = \tau\;
  {\rm Tr}\, \Big\{ v_\sigma \bar F^\sigma_{v'}\, \Gamma\, H_v \Big\} \,,
\end{equation}
for matrix elements between the states destroyed by the fields in $H_v$ and
$F_{v'}^\sigma$.  Here $\tau$ is a dimensionless function of $w$, and
$h_v^{(Q)}$ is the heavy quark field in the effective theory.  This matrix
element vanishes at zero recoil for any Dirac structure $\Gamma$ and for any
value of $\tau(1)$, since the $B$ meson and the $(D_1,\,D_2^*)$ mesons are in
different heavy quark spin symmetry multiplets, and the current at zero recoil
is related to the conserved charges of the spin-flavor symmetry. 
Eq.~(\ref{lo}) leads to the $m_{c,b}\to\infty$ predictions for the form factors
$f_i$ and $k_i$ given in Ref.~\cite{IWsr}.

At order $\Lambda_{\rm QCD}/m_{c,b}$ there are corrections originating from
the matching of the $b\to c$ flavor changing current onto the effective 
theory, and from order $\Lambda_{\rm QCD}/m_{c,b}$ corrections to the HQET
Lagrangian.  To leading order in $\alpha_s$, the current $\bar c\,\Gamma\,b$
is represented in HQET by 
\begin{equation}\label{current}
\bar c\, \Gamma\, b = \bar h_{v'}^{(c)}\, 
  \bigg( \Gamma - \frac i{2m_c} \overleftarrow D\!\!\!\!\slash\, \Gamma  
  + \frac i{2m_b}\, \Gamma \overrightarrow D\!\!\!\!\slash
  + \ldots \bigg)\, h_v^{(b)} \,.
\end{equation}
For matrix elements between the states destroyed by the fields in 
$F_{v'}^\sigma$ and $H_v$, the new order $\Lambda_{\rm QCD}/m_{c,b}$ 
operators in Eq.~(\ref{current}) are
\begin{eqnarray}\label{curr}
\bar h^{(c)}_{v'}\, i\overleftarrow D_{\!\lambda}\, \Gamma\, h^{(b)}_v &=&
  {\rm Tr}\, \Big\{ {\cal S}^{(c)}_{\sigma\lambda}\, 
  \bar F^\sigma_{v'}\, \Gamma\, H_v \Big\} \,, \nonumber\\*
\bar h^{(c)}_{v'}\, \Gamma\, i\overrightarrow D_{\!\lambda}\, h^{(b)}_v &=&
  {\rm Tr}\, \Big\{ {\cal S}^{(b)}_{\sigma\lambda}\, 
  \bar F^\sigma_{v'}\, \Gamma\, H_v \Big\} \,.
\end{eqnarray}
Unlike $B\to D^{(*)}$ decays, since the $(D_1,\,D_2^*)$ mesons and the $B$ 
are in different multiplets, there is no relation between ${\cal S}^{(c)}$ 
and ${\cal S}^{(b)}$.  The most general form for these quantities is
\begin{equation}\label{Sdef}
{\cal S}^{(Q)}_{\sigma\lambda} = v_\sigma \Big[ \tau_1^{(Q)}\, v_\lambda 
  + \tau_2^{(Q)}\, v'_\lambda + \tau_3^{(Q)}\, \gamma_\lambda \Big] 
  + \tau_4^{(Q)}\, g_{\sigma\lambda} \,.
\end{equation}
The functions $\tau_i$ depend on $w$, and have mass dimension one.  
They are not all independent.  The equation of motion for the heavy quarks, 
$(v\cdot D)\,h_v^{(Q)}=0$, implies
\begin{eqnarray}\label{const1}
w\,\tau_1^{(c)} + \tau_2^{(c)} - \tau_3^{(c)} &=& 0 \,, \nonumber\\*
  \tau_1^{(b)} + w\,\tau_2^{(b)} - \tau_3^{(b)} + \tau_4^{(b)} &=& 0 \,.
\end{eqnarray}
Using $i\partial_\nu\,(\bar h_{v'}^{(c)}\,\Gamma\,h_v^{(b)}) = 
(\bar\Lambda v_\nu-\bar\Lambda'v'_\nu)\,\bar h_{v'}^{(c)}\,\Gamma\,h_v^{(b)}$,
valid for matrix elements between the states in $F_{v'}^\sigma$ and in
$H_v$, together with the equation of motion for the heavy quarks, 
and the constraints in Eq.~(\ref{const1}), we obtain
\begin{eqnarray}\label{const2}
(w-1)\,(\tau_1^{(c)}-\tau_2^{(c)}) - \tau^{(c)}_4 
  &=& (w \bar\Lambda' - \bar\Lambda)\, \tau \,. \nonumber\\*
(w-1)\,(\tau_1^{(b)}-\tau_2^{(b)}) - \tau^{(b)}_4
  &=& (w\bar\Lambda - \bar\Lambda')\, \tau \,.
\end{eqnarray}
At zero recoil, 
$\tau^{(b)}_4(1)=-\tau^{(c)}_4(1)=(\bar\Lambda'-\bar\Lambda)\,\tau(1)$.  

Next we consider the terms originating from the order 
$\Lambda_{\rm QCD}/m_{c,b}$ corrections to the HQET Lagrangian, 
\begin{equation}\label{lag}
\delta {\cal L} = \frac1{2 m_Q}\, \bar h_{v}^{(Q)} \Big[ (iD)^2 
  + \frac{g_s}2\, \sigma_{\alpha\beta} G^{\alpha\beta}\, \Big] h_{v}^{(Q)} \,.
\end{equation}  
These corrections modify the heavy meson states compared to their infinite
heavy quark mass limit.  For example, they cause the mixing of the $D_1$ with
the $J^P=1^+$ member of the $s_l^{\pi_l}=\frac12^+$ doublet.  (This is a very
small effect, since the $D_1$ is not any broader than the $D_2^*$.)  The
kinetic energy operator does not violate heavy quark spin symmetry, and
therefore in $B\to D_1,D_2^*$ semileptonic decays its effect can be absorbed
into a redefinition of $\tau$, which is used hereafter.  For matrix elements
between the states destroyed by the fields in $F_{v'}^\sigma$ and $H_v$, the 
time ordered products of the chromomagnetic term in $\delta{\cal L}$ with the 
leading order currents are
\begin{eqnarray}\label{timeo}
i \int && {\rm d}^4x\, T\,\Big\{ \Big[ \bar h_{v'}^{(c)}\, 
  {g_s\over2}\,\sigma_{\alpha\beta}\, G^{\alpha\beta}\, h_{v'}^{(c)} 
  \Big](x)\, \Big[ \bar h_{v'}^{(c)}\, \Gamma\, h_{v}^{(b)} \Big](0)\, \Big\} 
  = {\rm Tr}\, \bigg\{ {\cal R}_{\sigma\alpha\beta}^{(c)}\,
  \bar F_{v'}^\sigma\, i\sigma^{\alpha\beta}\, \frac{1+\vslash'}2\, 
  \Gamma\, H_v \bigg\} \,, \nonumber\\*
i \int && {\rm d}^4x\, T\,\Big\{ \Big[ \bar h_v^{(b)}\, 
  {g_s\over2}\,\sigma_{\alpha\beta}\, G^{\alpha\beta}\, h_v^{(b)} \Big](x)\, 
  \Big[ \bar h_{v'}^{(c)}\, \Gamma\, h_{v}^{(b)} \Big](0)\, \Big\} 
  = {\rm Tr}\, \bigg\{ {\cal R}_{\sigma\alpha\beta}^{(b)}\,
  \bar F_{v'}^\sigma\, \Gamma\, \frac{1+\vslash}2\, i\sigma^{\alpha\beta}
  H_v \bigg\} \,. 
\end{eqnarray}
The most general parametrizations of ${\cal R}^{(c,b)}$ are
\begin{eqnarray}\label{Rdef}
{\cal R}_{\sigma\alpha\beta}^{(c)} &=& 
  \eta_1^{(c)}\, v_\sigma \gamma_\alpha \gamma_\beta 
  + \eta_2^{(c)}\, v_\sigma v_\alpha \gamma_\beta 
  + \eta_3^{(c)}\, g_{\sigma\alpha} v_\beta \,, \nonumber\\*
{\cal R}_{\sigma\alpha\beta}^{(b)} &=& 
  \eta_1^{(b)}\, v_\sigma \gamma_\alpha \gamma_\beta 
  + \eta_2^{(b)}\, v_\sigma v'_\alpha \gamma_\beta 
  + \eta_3^{(b)}\, g_{\sigma\alpha} v'_\beta \,.
\end{eqnarray}
Only the part of ${\cal R}_{\sigma\alpha\beta}^{(c,b)}$ antisymmetric in
$\alpha$ and $\beta$ contributes, when inserted into Eq.~(\ref{timeo}).  The
functions $\eta_i$ depend on $w$, and have mass dimension one.  (Note that
$g_{\sigma\alpha}\gamma_\beta$ is dependent on the tensor structures included
in Eq.~(\ref{Rdef}).)  All contributions arising from the time ordered products
in Eq.~(\ref{timeo}) vanish at zero recoil, since $v_\sigma\bar F_v^\sigma=0$,
and \hbox{$v_\alpha(1+\vslash)\sigma^{\alpha\beta}(1+\vslash)=0$}.%
\footnote{Order $\Lambda_{\rm QCD}/m_c$ corrections were also analyzed in
Ref.~\cite{Mannel}.  We find that $\tau_4$ (denoted $\xi_4$ in \cite{Mannel})
does contribute in Eq.~(\ref{curr}) for
$\Gamma=\gamma_\lambda\widetilde\Gamma$, and corrections to the Lagrangian are
parametrized by more functions than in \cite{Mannel}.}

Using Eqs.~(\ref{Hdef})--(\ref{Rdef}), it is straightforward to express the
form factors $f_i$ and $k_i$ parametrizing $B\to D_1\,\ell\,\bar\nu$ and $B\to
D_2^*\,\ell\,\bar\nu$ semileptonic decays in terms of Isgur-Wise functions.  
We use the constraints in Eqs.~(\ref{const1}) and (\ref{const2}) to eliminate
$\tau_3$ and $\tau_4$.  The form factors in Eq.~(\ref{formf}) depend on
$\tau_i^{(b)}$ and $\eta_i^{(b)}$ only through the linear combinations
$\tau_b=(2w+1)\,\tau_1^{(b)}+\tau_2^{(b)}+2(\bar\Lambda'-w\bar\Lambda)\,\tau$
and $\eta_b=6\,\eta_1^{(b)}-2(w-1)\,\eta_2^{(b)}+\eta_3^{(b)}$.  Denoting
$\varepsilon_{c,b}=1/(2m_{c,b})$ and dropping the superscript on $\tau_i^{(c)}$
and $\eta_i^{(c)}$, we obtain for the $B\to D_1\,\ell\,\bar\nu$ form factors
\begin{eqnarray}\label{expf}
\sqrt6\, f_A =&& - (w+1)\tau 
  - \varepsilon_b [ (w-1)\tau_b+(w+1)\eta_b ] \nonumber\\*
&& + \varepsilon_c [ 3(w-1) (\tau_1-\tau_2) 
  + (w+1) (2\eta_1+3\eta_3) 
  - 4(w\bar\Lambda'-\bar\Lambda)\tau ] ,\nonumber\\*
\sqrt6\, f_{V_1} =&&  (1-w^2)\tau 
  - \varepsilon_b (w^2-1) (\tau_b + \eta_b) \nonumber\\*
&& + \varepsilon_c [ (w^2-1)(3\tau_1-3\tau_2+2\eta_1+3\eta_3) 
  - 4(w+1)(w\bar\Lambda'-\bar\Lambda)\tau ] , \\
\sqrt6\, f_{V_2} =&&  -3\tau 
  - 3\varepsilon_b (\tau_b+\eta_b) 
  - \varepsilon_c [ (4w-1)\tau_1+5\tau_2 +10\eta_1 
  + 4(w-1)\eta_2-5\eta_3 ] ,\nonumber\\*
\sqrt6\, f_{V_3} =&&  (w-2)\tau 
  + \varepsilon_b [ (2+w)\tau_b - (2-w)\eta_b ] 
  + \varepsilon_c [ (2+w)\tau_1 + (2+3w)\tau_2 \nonumber\\* 
&&  - 2(6+w)\eta_1 - 4(w-1)\eta_2 - (3w-2)\eta_3
  + 4(w\bar\Lambda'-\bar\Lambda)\tau ] , \nonumber
\end{eqnarray}
The analogous formulae for $B\to D_2^*\,\ell\,\bar\nu$ are
\begin{eqnarray}\label{expk}
k_V &=& - \tau - \varepsilon_b (\tau_b+\eta_b)
  - \varepsilon_c (\tau_1-\tau_2-2\eta_1+\eta_3) , \nonumber\\*
k_{A_1} &=& - (1+w)\tau - \varepsilon_b [ (w-1)\tau_b+(1+w)\eta_b ] \nonumber\\*
&& - \varepsilon_c [ (w-1)(\tau_1-\tau_2)-(w+1)(2\eta_1-\eta_3) ] , \nonumber\\
k_{A_2} &=& - 2\varepsilon_c (\tau_1+\eta_2) , \\*
k_{A_3} &=& \tau + \varepsilon_b (\tau_b+\eta_b)
  - \varepsilon_c (\tau_1+\tau_2+2\eta_1-2\eta_2-\eta_3) . \nonumber
\end{eqnarray}

The allowed kinematic range for $B\to D_1\,\ell\,\bar\nu$ decay is $1<w<1.32$,
while for $B\to D_2^*\,\ell\,\bar\nu$ decay it is $1<w<1.31$.  Since these
ranges are fairly small, it is useful to expand the differential decay rates 
in Eq.~(\ref{rates}) simultaneously in powers of $w-1$ and $\Lambda_{\rm
QCD}/m_{c,b}$.
\begin{eqnarray}\label{exprates}
{{\rm d}\Gamma_{1,2} \over{\rm d}w} &\simeq& 
  {G_F^2\,|V_{cb}|^2\,m_B^5\,r_{1,2}^3 \over48\pi^3}\, \sqrt{w^2-1} \\*
&\times& \Big[ x_{1,2}^{(0)} + x_{1,2}^{(1)}\,(w-1) + x_{1,2}^{(2)}\,(w-1)^2 
  \Big]\, \tau^2(1) \,, \nonumber
\end{eqnarray}
In $x_{1,2}^{(i)}$ ($i=0,1,2$), we keep terms up to order 
$(\Lambda_{\rm QCD}/m_{c,b})^{2-i}$.  
Eqs.~(\ref{rates}), (\ref{expf}), and (\ref{expk}) yield
\begin{eqnarray}\label{x12}
x^{(0)}_1 &=& 32\,\varepsilon_c^2\, (\bar\Lambda'-\bar\Lambda)^2\, 
  (1-r_1)^2\, \,, \nonumber\\*
x^{(1)}_1 &=& (8/3)\, [(1-r_1)^2 + 4\varepsilon_c\, 
  (\bar\Lambda'-\bar\Lambda)\, (3-4r_1+r_1^2) \nonumber\\*
&& \phantom{(8/3)} - 2(1-r_1)^2\,(2\varepsilon_c\eta_1+3\varepsilon_c\eta_3
  -\varepsilon_b\eta_b ) / \tau ] \,, \nonumber\\*
x^{(2)}_1 &=& (8/3)\, [(3-4r_1+3r_1^2) + 2 (1-r_1)^2\, \tau'/\tau ] \,, 
  \nonumber\\
x^{(0)}_2 &=& 0 \,, \\*
x^{(1)}_2 &=& (40/3)\, (1-r_2)^2\, [1 - 2(2\varepsilon_c\eta_1 
  -\varepsilon_c\eta_3-\varepsilon_b\eta_b)/ \tau ] \,,\nonumber\\*
x^{(2)}_2 &=& 8\, (3-8r_2+3r_2^2) + (80/3)\, (1-r_2)^2\, \tau'/\tau \,. 
  \nonumber
\end{eqnarray}
Here $\tau$, $\tau'={\rm d}\tau/{\rm d}w$, and $\eta_i$ are evaluated at $w=1$.
The values of $\eta_i$ that occur in Eq.~(\ref{x12}) are not known.  Since the
$D_2^*-D_1$ mass splitting is very small, and model calculations indicate that
the analogous functions parametrizing time ordered products of the
chromomagnetic operator for $B\to D^{(*)}\,\ell\,\bar\nu$ decays are tiny
\cite{qcdsr}, hereafter we neglect the corrections parametrized by $\eta_i$.

The value of $\tau(1)$ can be determined from the experimental measurement of
the \hbox{$B\to D_1\,\ell\,\bar\nu$} branching ratio.  
We use the average of the ALEPH
\cite{ALEPH} and CLEO \cite{CLEO} results, ${\rm Br}(B\to
D_1\,\ell\,\bar\nu)=(6.1\pm1.1)\times10^{-3}$, to obtain 
\begin{equation}\label{tau1}
\tau(1)=0.55\pm0.05 \,.
\end{equation}
(We used $\bar\Lambda'-\bar\Lambda=0.35\,$GeV from Eq.~(\ref{LpL}),
$\tau_B=1.6\,$ps, $|V_{cb}|=0.04$, and $m_c=1.4\,$GeV.)  To get
Eq.~(\ref{tau1}) we assumed $\tau'(1)/\tau(1)=-0.8$.  $\tau(1)$
has little sensitivity to this choice.  Allowing $-1.2<\tau'(1)/\tau(1)<-0.5$
only affects the central value in Eq.~(\ref{tau1}) by $\pm0.01$.  The ISGW
nonrelativistic constituent quark model predicts $\tau(1)=0.54$ in surprising
agreement with Eq.~(\ref{tau1}) \cite{ISGW,IWsr}.

The ALEPH and CLEO analyses assume that $B\to D_1\,\ell\,\bar\nu\,X$ is
dominated by $B\to D_1\,\ell\,\bar\nu$, and that $D_1$ decays only into
$D^*\pi$.  If the first assumption turns out to be false then $\tau(1)$ will
decrease, if the second assumption is false then $\tau(1)$ will increase
compared to Eq.~(\ref{tau1}).  

Even though $\varepsilon_c(\bar\Lambda'-\bar\Lambda)\simeq0.12$ is small, the
term proportional to it comes with a large coefficient, and dominates the value
of $x^{(1)}_1$.  Numerically, this $\Lambda_{\rm QCD}/m_c$ correction to
$x_1^{(1)}$ is 1.8, while the part that survives in the $m_{c,b}\to\infty$
limit is 0.8.  Note that the part of the $\Lambda_{\rm QCD}^2/m_c^2$ correction
to $x_1^{(1)}$ that involves $\bar\Lambda'$, $\bar\Lambda$, and $\tau'(1)$ is
only $0.27\pm0.10$ for the previously mentioned range of $\tau'(1)$ (using
$\bar\Lambda=0.4\,$GeV \cite{gklw}).  The order $\Lambda_{\rm QCD}/m_c$ terms
that involve $\bar\Lambda'$ and $\bar\Lambda$ change $x_1^{(2)}$ by less than a
third of its leading order value for $-1.2<\tau'(1)/\tau(1)<-0.5$.  

After Eq.~(\ref{lag}), we absorbed into $\tau$ the form factor that
parametrizes time ordered products of the kinetic energy operator with the
leading order currents.  While $\lambda_1'$ is quite large (see
Eq.~(\ref{LpL})), this is probably a consequence of the $D_1$ and $D_2^*$ being
$p$-waves in the quark model, and does not necessarily imply that the heavy
quark kinetic energies significantly distort the overlap of wave functions that
yield the form factors.  If we had explicitly included the time ordered product
involving the charm quark kinetic energy, the leading term in $x^{(1)}_1$ would
change from $(1-r_1)^2$ to $(1-r_1)^2\,(1+2\eta_{\rm ke}\,\varepsilon_c)$. 
Even taking $\eta_{\rm ke}=\pm\bar\Lambda'$ changes the extracted value of
$\tau(1)$ by less than 0.04.  So this $\Lambda_{\rm QCD}/m_c$ correction to
$x^{(1)}_1$ is likely to be much smaller than the term proportional to
$\varepsilon_c(\bar\Lambda'-\bar\Lambda)$ explicitly shown in Eq.~(\ref{x12}). 
It is important to have experimental data on the $w$-spectrum of $B\to
D_1\,\ell\,\bar\nu$ decay to test the hypothesis that the effects of the
kinetic energy and $\Lambda_{\rm QCD}^2/m_c^2$ corrections are not large.  

The order $\Lambda_{\rm QCD}/m_{c,b}$ corrections calculated in this letter are
also important for the prediction of $R\equiv{\rm Br}(B\to
D_2^*\,\ell\,\bar\nu)/{\rm Br}(B\to D_1\,\ell\,\bar\nu)$.  As $R$ is sensitive
to $\tau'(1)/\tau(1)$, we shall explicitly display the dependence on
$\tau'(1)$.  In the $m_{c,b}\to\infty$ limit, expanding to linear order in
$\tau'(1)/\tau(1)$, we obtain $R=1.89+0.51\,\tau'(1)/\tau(1)$.  Including the
$\Lambda_{\rm QCD}^2/m_c^2$ correction to $x^{(0)}_1$, and the $\Lambda_{\rm
QCD}/m_c$ correction to $x^{(1)}_1$, and expanding again to linear order in
$\tau'(1)/\tau(1)$, yields $R=0.79+0.30\,\tau'(1)/\tau(1)$.  This suppression
of $R$ compared to the infinite mass limit is consistent with experimental
data.  (It is possible that part of the reason for ${\rm Br}(B\to
D_2^*\,\ell\,\bar\nu\,X)\times{\rm Br}(D_2^*\to
D^{(*)}\pi)\lesssim(1.5-2.0)\times10^{-3}$ \cite{ALEPH} is a suppression of
${\rm Br}(D_2^*\to D^{(*)}\pi)$ compared to ${\rm Br}(D_1\to D^*\pi)$.)

Our results are important for sum rules that relate inclusive $B\to
X_c\,\ell\,\bar\nu$ decays to the sum of exclusive channels.  The Bjorken sum
rule \cite{Bjsr,IWsr} for the slope of the $B\to D^{(*)}\,\ell\,\bar\nu$
Isgur-Wise function becomes $\rho^2\equiv-{\rm d}\xi/{\rm
d}w|_{w=1}>0.25+\tau^2(1)=0.55$.  Note that $2\tau^2(1)/3$ arises from the
$D_1,\,D_2^*$ doublet, while $\tau^2(1)/3$ is due to the broad
$s_l^{\pi_l}=\frac12^+$ doublet, ($D_0^*,\,D_1^*$).  (Using the equality of the
leading Isgur-Wise functions for these multiplets in the quark model, valid for
any spin-orbit independent potential.)
A class of zero recoil sum rules were considered in Ref.~\cite{VVsr}.  The
axial sum rule, which bounds the $B\to D^*$ form factor that measures
$|V_{cb}|$, receives no corrections from either the $\frac32^+$ or the
$\frac12^+$ doublets.  The $J^P=1^+$ states contribute to the vector sum rule,
which bounds the $\lambda_1$ parameter of HQET.  This bound is strongest in the
limit $m_c\gg m_b\gg\Lambda_{\rm QCD}$, where the $D_1$ state does not
contribute.  The equality of Isgur-Wise functions for the $\frac32^+$ and
$\frac12^+$ doublets in the quark model implies that $B\to D_1^*$ transition
modifies the bound to
$\lambda_1<-3\lambda_2-3(\bar\Lambda^*-\bar\Lambda)^2\,\tau^2(1)
\simeq-3\lambda_2-0.09$.  (Here $\bar\Lambda^*\simeq\bar\Lambda+0.31\,$GeV
\cite{GI} is the analogue of $\bar\Lambda$ for the $\frac12^+$ states.)
Perturbative corrections to these bounds can be found in \cite{pert}.

In this letter we analyzed $B\to D_1(2420)\,\ell\,\bar\nu$ and $B\to
D_2^*(2460)\,\ell\,\bar\nu$ decay form factors at order $\Lambda_{\rm
QCD}/m_{c,b}$.  At zero recoil, all $\Lambda_{\rm QCD}/m_{c,b}$ corrections can
be written in terms of the $m_{c,b}\to\infty$ Isgur-Wise function for these
transitions, and known meson mass splittings.  With some model dependent
assumptions, we predicted the shape of the spectrum near zero recoil, including
order $\Lambda_{\rm QCD}/m_{c,b}$ corrections.  Testing these predictions will
constitute an interesting check on our understanding of exclusive semileptonic
decays based on the HQET.  Similar results hold for semileptonic $B$ decay into
the broad $s_l^{\pi_l}=\frac12^+$ charmed meson multiplet, and for the
semileptonic $\Lambda_b$ decays into excited charmed baryons.  These will be
presented in a separate publication.  Perturbative QCD corrections and
nonleptonic decays (using factorization) will also be considered there.

\acknowledgments
We thank Isi Dunietz for helpful discussions.  
This work was supported in part by the Department of Energy
under grant no.~DE-FG03-92-ER 40701.

{\tighten

} %end tighten (references & figure captions)


\begin{references}


\bibitem{HQS}
N. Isgur and M.B. Wise, Phys. Lett. B232 (1989) 113; 
Phys. Lett. B237 (1990) 527.

\bibitem{NuWe}
S. Nussinov and W. Wetzel, Phys. Rev. D36 (1987) 130.

\bibitem{VoSi}
M. Voloshin and M. Shifman, Sov. J. Nucl. Phys. 47 (1988) 511.

\bibitem{Luke}
M.E. Luke, Phys. Lett. B252 (1990) 447.

\bibitem{eft}
E. Eichten and B. Hill, Phys. Lett. B234 (1990) 511;
H.~Georgi, Phys. Lett. B240 (1990) 447.

\bibitem{IWprl}
N. Isgur and M.B. Wise, Phys. Rev. Lett. 66 (1991) 1130.

\bibitem{ALEPH}
ALEPH Collaboration, Report no.~PA01-073, ICHEP96 Conference, 
25-31 July 1996, Warsaw, Poland.

\bibitem{CLEO}
CLEO Collaboration, T.E. Browder {\it et al.}, Report no. CLEO CONF 96-2,
ICHEP96 PA05-077.

\bibitem{OPAL}
OPAL Collaboration, R. Akers {\it et al.}, Z. Phys. C67 (1995) 57.

\bibitem{PDG}
Particle Data Group, R.M. Barnett {\it et al.}, Phys. Rev. D54 (1996) 1.

\bibitem{trace}
A.F. Falk {\it et al.}, Nucl. Phys. B343 (1990) 1; \\
J.D. Bjorken, {\it Proceedings of the 18th SLAC Summer Institute on
Particle Physics}, pp.\ 167, Stanford, July 1990, ed.~by
J.F. Hawthorne (SLAC, Stanford, 1991); \\
A.F. Falk, Nucl. Phys. B378 (1992) 79.

\bibitem{IWsr}
N. Isgur and M.B. Wise, Phys. Rev. D43 (1991) 819.

\bibitem{Mannel}
T. Mannel and W. Roberts, Z. Phys. C61 (1994) 293.

\bibitem{qcdsr}
M. Neubert {\it et al.}, Phys. Lett. B301 (1993) 101; 
Phys. Rev. D47 (1993) 5060.

\bibitem {ISGW}
N. Isgur {\it et al.}, Phys. Rev. D39 (1989) 799.

\bibitem{gklw}
M. Gremm {\it et al.}, Phys. Rev. Lett. 77 (1996) 20.

\bibitem{Bjsr}
J.D. Bjorken, Invited talk at Les Rencontre de la Valle d'Aoste
(La Thuile, Italy), SLAC-PUB-5278 (1990).

\bibitem{VVsr}
I.I. Bigi {\it et al.}, Phys. Rev. D52 (1995) 196.

\bibitem{GI}
S. Godfrey and N. Isgur, Phys. Rev. D32 (1985) 189.

\bibitem{pert}
A. Kapustin {\it et al.}, Phys. Lett. B375 (1996) 327; \\
C.G. Boyd {\it et al.}, Phys. Rev. D55 (1997) 3027.


\end{references}
\end{document}